\documentclass[prb,aps,twocolumn,showpacs,amsmath,amssymb]{revtex4}
\usepackage{epsf}

\begin{document}
\draft
\newcommand{\ve}[1]{\boldsymbol{#1}}

\title{Modulation of Superconducting Properties by Ferroelectric Polarization in Confined FE-S-FE Films}
\author{Natalia Pavlenko}
\address{Institute of Physics, University of Augsburg, 86135 Augsburg, Germany\\
Institut f\"{u}r Theoretische Physik T34, Physik-Department
der TU M\"{u}nchen, James-Franck-Strasse, 85747 Garching
b.~M\"{u}nchen, Germany}
\date{\today}

\begin{abstract}
We show that the electric polarization at the interface with
ultrathin superconducting (S) films sandwiched
between ferroelectric (FE) layers allows achievement of substantially stronger
modulation of inner carrier density and superconducting transition
temperature as compared to FE-S bilayers typically used in
superconducting FETs. We find that not only the larger penetration depths but also
the pairing symmetry should be responsible for the fact that the electric field effect
in high temperature superconductors is much stronger than in conventional systems.
Discussing the advantages of multilayers, we
propose a novel design concept for superconducting electric
field-effect transistors based on ferroelectric films.
\end{abstract}


\maketitle
\section{Introduction}
Electric field effect in superconductors continues to attract considerable
attention in science and technology \cite{ahn}. The external field can modulate
the charge density and resistance, and control a reversible
superconductor-insulator switching behavior which plays a key role in the
development of superconducting field-effect transistors (SuFETs)
\cite{watanabe}. Especially in the high-T$_c$ superconducting cuprates, the
field effect is expected to be strong, since their low carrier density leads to
larger electric field penetration depths $\sim 0.5-1$~nm. Recently,
complex ferroelectric oxide materials with high dielectric constant like
Ba$_x$Sr$_{1-x}$TiO$_3$ (BST) and Pb(Zr$_x$Ti$_{1-x}$)O$_3$ (PZT)\cite{ahn2} have been
used to achieve substantial carrier modulation and shift of $T_c^S$ of several K.

Due to the small width of the accumulation layers at the contacts, the effect
of the polarization-induced field is most significant in ultrathin
superconducting films of a few nanometers thickness\cite{ahn,xi,ahn2}. With the
difficulties related to the fabrication of ultrathin films and interfaces of
good quality, theoretical modelling can be an additional effective tool to
study these systems. In the bilayer structures, the ferroelectric polarization
at the interface attracts or repels the charge carriers in the superconducting
film, in close analogy to the effect of doping
\cite{ahn2,watanabe,logvenov}. However, the field effect in FE-S {\it
multilayers} where the polarization acts on both superconducting
surfaces and may strongly modify the internal charge distribution in the entire
ultrathin S-film, is a nontrivial and challenging problem. Motivated by
recent experiments\cite{logvenov}, in this work we show that the use of
multilayers can drastically amplify the field effect, hence leading to much
stronger increase of the superconducting transition temperature in comparison
to S-FE bilayers typically exploited in SuFETs.

\section{The Model}
\begin{figure}[htbp]
\epsfxsize=6.5cm \centerline{\epsffile{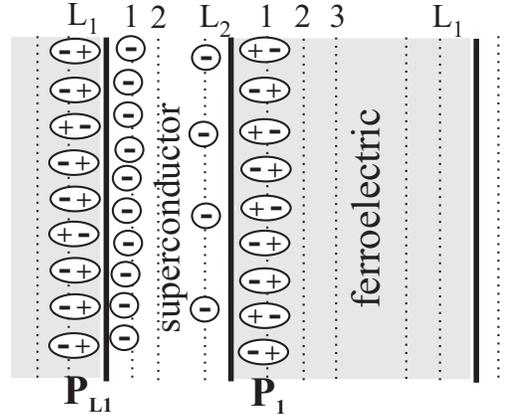}} \caption{Scheme of the
periodic FE-S multi-layer structure.} \label{fig1}
\end{figure}

In the approach considered here, we directly describe the internal charge redistribution
in a ultrathin S-film caused by the ferroelectric polarization. Assuming the existence
of an effectice attractive interaction, we focus on the question of how the
superconducting state in the film can be influenced by this polarization.
Specifically, we consider a hybrid system composed of periodically
alternating FE- and S-layers with
the ferroelectric polarization $\ve{P}$ directed perpendicular to
the interface, as shown in Fig.~\ref{fig1}. The thickness of the FE subsystem is given by the
number $L_1$ of monolayers $i_f=1, \ldots, L_1$
in each ferroelectric layer. In the S-film, $L_2$ denotes the number of
superconducting planes. We propose to describe the structure by the following model:\\
(i) The two possible orientations of ferroelectric dipoles (shown in
Fig.~\ref{fig1}) caused by the ion displacements on each lattice site (unit
cell) $l=1,\ldots, N_{\perp}$ of the $i_f$-th monolayer in the FE-layer can be
described by two values $\pm 1/2$ of a pseudospin operator $s(i_f,l)$. The
spontaneous polarization below the ferroelectric Curie temperature $T_c^F$ can
then be modelled by an Ising-type Hamiltonian on the cubic lattice \cite{lines}:
$H_F=-J_F\sum s(i_f,l) s(i_f',l')$, where the summation ranges over all nearest
neighboring sites; $J_F$ denotes the internal dipolar interaction potential.
The local polarization is given by the thermal average of $s_{i_f,l}$ and is
assumed to depend only on the monolayer index $i_f$: $\langle s_{i_f,l}
\rangle=\langle s_{i_f}
\rangle=P_{i_f}$.\\
(ii) In the superconducting film, we consider a
BCS-like pairing in the planes $i_s=1, \ldots, L_2$:
\begin{eqnarray}\label{H_S}
H^{i_s}_S=\sum_{\ve{k},\sigma} (\varepsilon^{i_s}_{\ve{k}}-\mu)
n_{i_s,\ve{k},\sigma}-\sum_{\ve{k}}(\Delta^{i_s}
b^{\dag}_{i_s,\ve{k}}+h.c.+ C_{\ve{k}})
\end{eqnarray}
The planes are coupled via the interplanar tunneling
of the bosonlike pairs,
\begin{eqnarray}\label{H_t}
H_{\perp}=-t_{\perp} \sum_{i_s,\ve{k}} (b_{i_s,\ve{k}}^{\dag}
b_{i_s+1,\ve{k}}+h.c.)
\end{eqnarray}
where the pair operators are
$b_{i_s,\ve{k}}=c_{i_s,-\ve{k},\downarrow}c_{i_s,\ve{k},\uparrow}$, the
operator $c^{\dag}_{i_s,\ve{k},\sigma}$ creates an electron with spin $\sigma$
and impulse $\ve{k}$ in the $i_s$-th plane, and
$C_{\ve{k}}=\Delta^{i_s}_{\ve{k}} \langle b_{i_s,\ve{k}}^{\dag} \rangle$. In
(\ref{H_S}), $\varepsilon^{i_s}_{\ve{k}}=-2 t (\cos k_x + \cos k_y)$ refers to
the electron band energy with lattice constant $a=1$, and $\mu$ is the chemical
potential. Assuming that some amount of charge
$N=\sum_{i_s,\ve{k},\sigma} \langle n_{i_s,\ve{k},\sigma} \rangle$ is either
injected into the S-film from the electrodes or already exists due to
chemical doping, we concentrate here {\it on the effect of the carrier
redistribution between the S-planes} due to the contact polarization. For $s$-wave pairing,
the superconducting pairing amplitude in each plane
is: $\Delta^{i_s}=\Delta^{i_s}_{\ve{k}}=-\sum_{\ve{k'}} V^{i_s}_{\ve{k}\ve{k'}}
\langle b_{i_s,\ve{k'}}\rangle$, with the
effective pairing potential $V^{i_s}_{\ve{k}\ve{k'}}=-V^{i_s}$.\\
(iii) At the interface we describe the screening of the surface charge due
to the FE-polarization on the surfaces $i_f=1$ and $i_f=L_1$ by the electrons of the S-film.
We consider here only the electrostatic interaction with the electron density
of the boundary S-planes $i_s=1$ and $i_s=L_2$ given by the
electron number operators $n_{i_s,l,\sigma}$ \cite{pavlenko,pavlenko2},
\begin{equation}\label{H_int}
H_{int}=\gamma \sum_{l,\sigma} s_{1,l}\cdot n_{L_2,l,\sigma}-\gamma
\sum_{l,\sigma} s_{L_1,l}\cdot n_{1,l,\sigma},
\end{equation}
where $\gamma$ is the charge-ferroelectric dipole
interaction energy. Note that, as the screening length in the cuprate superconductors is about 1~nm,
the S-planes close to the boundary S-planes should also be taken into account in (\ref{H_int})
in the multilayered cuprates containing more than one CuO$_2$ plane in the unit cell\cite{kotegawa}
with a distance $\sim 3.2$~$\AA$ between the planes.

Furthermore, we focus here on the case where the FE-layers are far below the
Curie temperature $T_c^F$. For this case, we neglect the polarization
fluctuations and rewrite the energy
$H_F+H_{int}$ in the mean-field approximation, whereas the quasiparticle
energies for each given configuration $|(1,\ve{k}_1,\uparrow),
(1,-\ve{k}_1,\downarrow), \ldots, (L_2,\ve{k}_{L_2},\uparrow),
(L_2,-\ve{k}_{L_2},\downarrow) \rangle$ are found by exact numerical
diagonalization of the effective S-film energy $\left \{
\sum_{i_s}H_S^{i_s}+H_{\perp}+H_F^{MF}+H_{int}^{MF}\right \}$. The
corresponding set of the order parameters $\Delta^{i_s}$ and $P_{i_f}$ is
calculated selfconsistently by minimization of the system free energy. As a
result, the proposed procedure describes the inner charge redistribution
between the planes of an S-film in effective polarized
medium on the boundaries.
\begin{figure}[htbp]
\epsfxsize=7.5cm \centerline{\epsffile{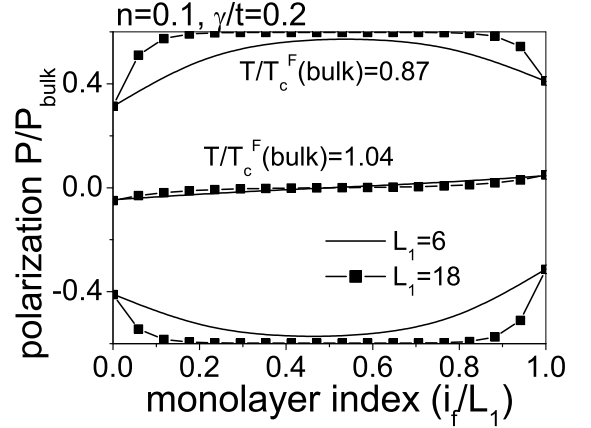}} \caption{Temperature
evolution of polarization profiles in a heterostructure containing $L_2=2$
S-monolayers for $J_F/t=1$.} \label{fig2}
\end{figure}

The electron-dipole interaction $\gamma=eZ u_{FE}/d_{SF}^2$ ($e$ and $Z$
are the electronic and the net FE-unit cell charges) depends on the
distance $d_{SF}$ between the nearest FE-monolayers and S-planes. It depends also on the
ionic displacement amplitude $u_{FE}$ in the FE-unit cell. For instance, at
$d_{SF}=5$~\AA, for a BST-layer with polarization $P_s \sim
26$~$\mu$C$/cm^2$ and $u_{FE} \sim 0.1$~\AA, we obtain $\gamma \sim
0.03$~eV, whereas for $u_{FE} \sim 0.5$~\AA (like in LiNbO$_3$) we get $\gamma
\sim 0.15$~eV. A comparison with the case of a FE-S bilayer
is easily provided by fixing the coupling $\gamma$ at
one of the contacts to zero. In our calculations, we take the pairing potential
$V_0=V^{i_s}=3.5t$ independent of $i_s$. Considering the ideal surfaces, we should note
that in SuFETs the surface roughness leads to the interface steps with a height
$h \sim 1$ unit cell of SrTiO$_3$ as reported in Ref.~\onlinecite{logvenov}. In the ultrathin S-films,
such a step can act as a weak link and strongly affect the in-plane transport \cite{wehrli}.
As long as $h$ does not exceed the interplanar distance in the S-film, we can expect
our results to not be significantly affected by the rough interfaces, but the problem of
transport in this case needs more careful analysis.

\section{Ferroelectric Polarization}
The selfconsistent solutions show stabilization of two different ferroelectric
phases (+) and ($-$) (discussed in details in Ref.~\onlinecite{pavlenko}) with
decreasing temperature $T$, depending on the electron concentration
$n=N/(N_{\perp} L_2)$ and $\gamma$. In phase (+) (Fig.~\ref{fig2}), the
electric dipoles have the same orientation in each FE-layer (as shown in the
inset of Fig.~\ref{fig3}(a)), and the substantial deviations of the two
possible solutions for $P$ ($P_{i_f}^1=-P_{L_1-i_f+1}^2$) due to the electron
screening (\ref{H_int}) appear close to the interfaces and decay exponentially
beyond the distance of about the ferroelectric correlation
length\cite{pavlenko2}. As seen in Fig.~\ref{fig2}, the polarization is also
suppressed for smaller FE-layer thickness due to the depolarization field
$E_d=-J_F\sum(\langle s(i_f',l') \rangle-P_{bulk})$ acting on each dipole
$s(i_f,l)$, however the suppression obtained here is not as crucial as in
Ref.~\onlinecite{ghosez} and we still find a nonzero $P$ for a layer with
$L_1=6$ monolayers. To get stronger suppression for a concrete FE-compound, one
should consider more realistic long-range dipole-dipole interactions
\cite{lines} as well as strain effects. Nevertheless, we expect our main
results related to the superconducting properties robust, since we consider
the coupling $\gamma$ in the range $\gamma/t<1$ ($t\sim 0.1$~eV for
superconducting cuprates) which corresponds to the polarization $P_s <
25$~$\mu$C$/cm^2$ obtained in Ref.~\onlinecite{ghosez} for FE-films of about
10~nm thickness. Also, taking into account the $P$-fluctuations, would suppress
the mean-field values of $P$ considered in this work as an upper boundary for
polarization profiles. With increasing coupling $\gamma$ ($\gamma/t>1$) and for
higher electron densities $n>0.5$, a transition into the
ferroelectric phase ($-$) occurs. The fluctuating polarization domains\cite{pavlenko}
appearing in the phase ($-$), are a property of the Ising model in thin films.
However, in a specific ferroelectric material these domains could be suppressed
due to stronger depolarization fields \cite{domains}. As the modulation of
charge density in this state is much weaker than in phase (+), we focus here
only on the polarization in the phase (+) which is of crucial importance
for the superconducting properties.

\section{Redistribution of Charge in Superconducting Film}
\begin{figure}[htbp]
\epsfxsize=8.5cm {\epsffile{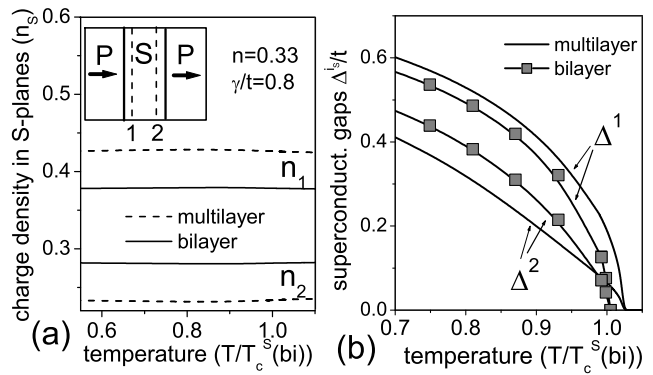}} \caption{(a) Charge densities vs $T$ in
the S-planes $i_s=1,2$ in the multi- and bilayer systems containing $L_1=10$
FE-monolayers in the phase (+) with $J_F/t=1$, $t_{\perp}/t=0.1$. (b) The corresponding pairing
amplitudes $\Delta^{i_s}$ are scaled by $t$.} \label{fig3}
\end{figure}

In the phase (+), the polarization $P^1>0$ attracts
the charge in the plane $i_s=1$ ($\Delta n_1=n_1-n>0$) while
$n_{L_2}$ decreases ($\Delta n_{L_2}=n_{L_2}-n<0$) as illustrated in Fig.~\ref{fig3}(a).
At low temperatures, the charge density redistribution
due to $P\ne 0$, can be described by the difference
$\Delta n=n_1-n_{L_2}\approx \frac{\gamma}{4t}(P_1+P_{L_1})>0$.
Consequently, the solution $P^2<0$ leads to $\Delta n<0$. Moreover,
in contrast to the bilayer system, the
same direction of $P$ in the second FE-layer, say, $P>0$, repels the charge near the right
contact and pushes it towards the accumulation region at the left
contact, acting as {\it an additional driving force} for the charge
supply into the accumulation plane $i_s=1$. This results in $n_1(multi)>n_1(bi)$ and
$\Delta n(multi)>\Delta n(bi)$ which
is clearly seen in Fig.~\ref{fig3}(a).

\begin{figure}[htbp]
\epsfxsize=7.5cm {\epsffile{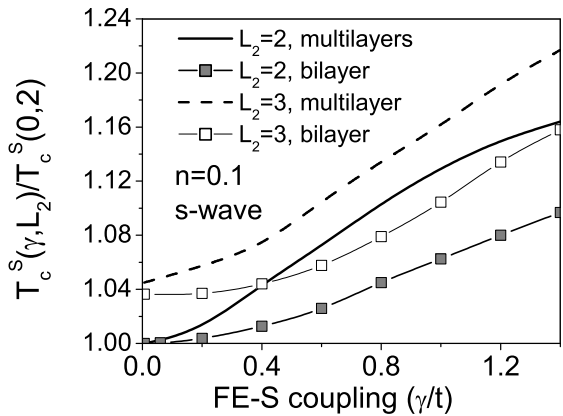}} \caption{$T_c^S$ vs coupling $\gamma$ in
a system with $L_1=10$ FE-monolayers for $n=0.1$ and $J_F/t=1$.} \label{fig4}
\end{figure}

To understand the effect of the inter-planar charge redistribution on the superconducting
transition temperature $T_c^S$, we consider the case $L_2=2$. With
the interplanar tunneling $t_{\perp}\rightarrow 0$, the Hamiltonian (\ref{H_S})-(\ref{H_int}) leads
to two gap equations in the planes $i_s$.
For $s$-wave symmetry, different local $T_c^S$ are found:
\begin{eqnarray}\label{tc_local}
T_c^{S(1)}\propto 2t\exp (-8t/V_0+\delta), T_c^{S(2)}\propto 2t\exp (-8t/V_0-\delta),
\end{eqnarray}
due to the factor $\delta=\frac{\gamma(1-n)}{8tn(2-n)} (P_1+P_2)$ caused by the
contact polarization.
Hence, for low band filling ($n<1$) and for $P>0$, one finds that $T_c^{S(1)}>T_c^{S(2)}$ and
$\Delta^1>\Delta^2$, since the gaps $\Delta^{i_s}(T=0)$ are comparable in energy to
$T_c^{S(i_s)}$.
For $T_c^{S(2)}<T<T_c^{S(1)}$, this leads to a normal state
in the plane $i_s=2$, whereas the plane $i_s=1$ is still superconducting. However,
due to the interplanar coupling, even slightly above $T_c^{S(2)}$,
the superconducting plane $i_s=1$ still induces a nonzero $\Delta^2 \ne 0$
for $T<T_c^{S(1)}$ stabilizing the superconductivity in the entire S-film.
This effect can be seen in Fig.~\ref{fig3}(b) from the numerical solution of the gap
equations for $t_{\perp}/t=0.1$.
For $t_{\perp}\ne 0$, both gaps $\Delta^1$ and $\Delta^2$ vanish at a common $T=T_c^S$ which is
higher than the local $T_c^{S(2)}$ calculated for $t_{\perp}=0$. The enhancement of the
pairing on the planes without the pairing interaction solely due to the interplanar coupling is
a well known property discussed in the literature\cite{donovan}.
In our case, however, the interplanar
tunneling plays not only the role of an enhancement factor, but also provides a way for the
accumulation of charge and stronger pairing in the accumulation S-plane caused by the polarization.
In the multilayers, due to the stronger redistribution of the charge described by $\Delta n$,
the last property leads to
$\Delta^1(multi)>\Delta^1(bi)$ and $T_c^S(multi)>T_c^S(bi)$ (Fig.~\ref{fig3}(b)).
For stronger coupling, the increase of
$T_c^S$ in the multilayer is also substantially stronger than
that in the bilayer. The strongest increase of $T_c^S$ caused by $P$ can be achieved
for low $n$; compared to the case $\gamma=0$ (isolated the S-layer),
we find a $25-35\%$ increase of $T_c^S$ with $\gamma/t\sim 1$ for $n=0.1$, see Fig.~\ref{fig4}.

\section{Inter-planar coupling}
To estimate the effect of $P$ for
$t_{\perp} \ne 0$, we rewrite the interplanar coupling in (\ref{H_t}) in the mean-field approximation
and for $L_2=2$ introduce the order parameters as
\begin{eqnarray}\label{delta_t}
{N_{\perp}}\Delta_{\ve{k}}^1=-\sum_{\ve{k'}} V_{\ve{k}\ve{'}}^{1} \langle b_{1,\ve{k'}} \rangle+
\sum_{\ve{k'}} t_{\perp}^{\ve{k}\ve{k'}} \langle b_{2,\ve{k'}} \rangle,\\
{N_{\perp}}\Delta_{\ve{k}}^2=-\sum_{\ve{k'}} V_{\ve{k}\ve{k'}}^{2} \langle b_{2,\ve{k'}} \rangle+
\sum_{\ve{k'}} t_{\perp}^{\ve{k}\ve{k'}} \langle b_{1,\ve{k'}} \rangle\nonumber
\end{eqnarray}
where the susceptibilities $\langle b_{i_s,\ve{k}}
\rangle=\frac{1}{2}\frac{\Delta_{\ve{k}}^{i_s}} {E_{\ve{k}}^{i_s}} \tanh
\left(\frac{1}{2}\beta E_{\ve{k}}^{i_s} \right)$,
$E_{\ve{k}}^{i_s}$ are
the eigenvalues of the mean-field superconducting Hamiltonian, and $\beta=\frac{1}{k_B T}$.
The inter-layer tunneling $t_{\perp}^{\ve{k}\ve{k'}}$ is introduced in a more
generalized form so that the coherent tunneling (\ref{H_t}) corresponds to
$t_{\perp}=t_{\perp}^{\ve{k}\ve{k'}}\delta_{\ve{k}\ve{k'}}$. As the field
effect is expected to be stronger in high-T$_c$ superconductors, we consider
further the two cases of $s$- and $d$-wave pairing with the ansatz
$\Delta_{\ve{k}}^{i_s}= \Delta^{i_s} \eta_{\ve{k}}$, $V_{\ve{k}\ve{k'}}=-V^{0}
\eta_{\ve{k}}\eta_{\ve{k'}}$ and $t_{\perp}^{\ve{k}\ve{k'}}=t_{\perp}
\eta_{\ve{k}}\eta_{\ve{k'}}$ in (\ref{delta_t}). Here $\eta_{\ve{k}}=1$ for 
the $s$-wave and $\eta_{\ve{k}}=\cos(k_x)-\cos(k_y)$ for
the $d$-wave pairing symmetry. As for $P>0$ we have
$\Delta^1>\Delta^2$ in the case of isolated S-planes, we assume that the second
term in the r.h.s. of the second equation (\ref{delta_t}) determining
$\Delta^2$ is dominant and substitute it into the first equation for
$\Delta^1$. The resulting quadratic equation has the
following solution for small $t_{\perp}/V^0 \ll
1$:
\begin{eqnarray}\label{tc_global}
T_c^{S}\propto 2t\exp (-8t/\tilde{V}^0+\tilde{\delta})),
\end{eqnarray}
which corresponds in fact to the upper local transition temperature given in
(\ref{tc_local}) in the limit $t_{\perp}\rightarrow 0$. However, in contrast to
the expressions (\ref{tc_local}), here due to the
inter-planar coupling we note three important features: (i) as
analyzed in the literature, the pairing interaction is increased:
$\tilde{V}^0=V^0(1+(\frac{t_{\perp}}{V^0})^2)$; (ii) the factor
$\tilde{\delta}=\frac{|\delta|}{1+2(\frac{t_{\perp}}{{V^0}})^2}$ always leads
to an increase of $T_c^S$ and does not depend on the direction of $P$,
demonstrating that the enhancement of $T_c^S$ is essentially due to the S-planes with
the dominant pairing (those with the higher local $T_c^{S(i_s)}$ in
(\ref{tc_local})); (iii) interestingly, the stronger inter-planar coupling
makes the field-effect slightly weaker, since the contribution of
$\tilde{\delta}$ in (\ref{tc_global}) decreases with the $t_{\perp}$ increase.
\begin{figure}[htbp]
\epsfxsize=7.5cm {\epsffile{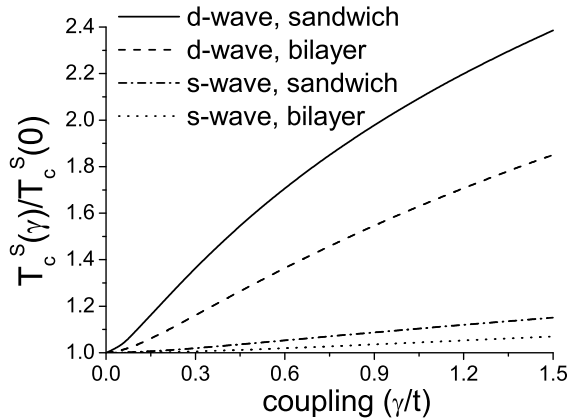}}\caption{$T_c^S$ versus $\gamma$ in P-S-P
sandwiches and S-P bilayers for the cases of s- and d-wave symmetries
calculated from (\ref{delta_t}). Here $n=0.3$, $t_{\perp}/t=0.05$, and the
contact polarization $P_1=P_{L_1}=\tanh \left(\frac{1}{2}\beta E_g \right)$ fixed by an
external field $E_g/t=0.5$.} \label{fig5}
\end{figure}
The same effect, i.e. the slightly weaker increase of $T_c^S$ for larger $t_{\perp}$ is also seen in
the numerical solution of (\ref{delta_t}). The most striking fact, however, is that
the coupling with the polarization leads to a much stronger increase of $T_c^S$ in the case of
$d$-wave pairing when compared to $s$-wave, shown in Fig.~\ref{fig5}. This result suggests that
{\it not only the larger screening length, but also the pairing symmetry should be responsible for
the much stronger electric field effect in the high-temperature superconductors} in comparison with the conventional
compounds. Also, the difference between the $T_c^S$ in the bilayers and sandwiches
for the $d$-wave symmetry is much more pronounced which allows us to expect further progress
in the field-effect experiments performed with FE-S-FE sandwiches.

\section{Scheme of sandwich-based SuFET}
\begin{figure}[htbp]
\epsfxsize=7.5cm {\epsffile{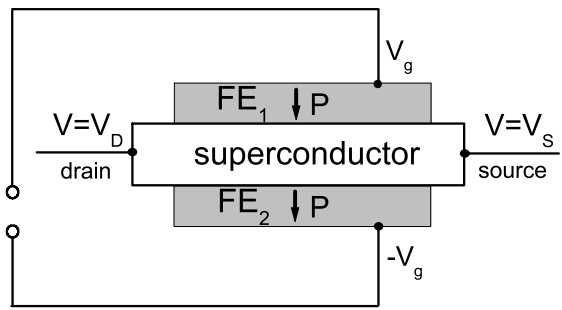}}\caption{Proposed scheme of SuFET based
on confined FE-S-FE geometry.} \label{fig6}
\end{figure}

Having demonstrated the advantages of sandwiches, we propose a design of a
field-effect transistor based on the confined FE-S-FE geometry.
The power supply moves the opposite charge to the FE-gate electrodes. The operational principle of
such an SuFET consists of two stages: (i) the superconducting state with the
enhanced $T_c^S$ can be reached in the regime where the system
has the same direction of $P$ in each FE-layer, as
illustrated in Fig.~\ref{fig6}. Note that for the fixed $n$, the antiparallel polarization
in the FE-gates
leads to lower charge densities in the accumulation region and consequently to the lower $T_c$
which is discussed in Ref.~\onlinecite{pavlenko3}. (ii) To switch the SuFET into the insulating
state, it is sufficient to destroy the accumulation layer at the first FE-S
contact. This can be realized by switching the voltage $V_g \rightarrow -V_g$
and tuning the polarization in the FE-layers to zero.
Due to the simple geometry, even at the current stage of  fabrication
techniques, the proposed SuFETs can be designed and experimentally
probed with oxide films as the most promising candidates for
high-temperature superconducting electronics. From the above results we expect that
in the proposed SuFETs also the effect of charge tunneling between the CuO$_2$-planes could be tested
in single- and especially in multilayered cuprates with inhomogeneous carrier
distribution\cite{kotegawa}.

In conclusions, we have shown that the FE-polarization in
FE-S multilayers leads to much stronger modulation of
inner carrier density and superconducting properties as
compared to the bilayers. These advantages of
multilayers can be used as the basis of novel design of proposed
SUFETs.

\section*{Acknowledgements}
This work has been supported by the DFG Grants No.~SPP-1056 and SFB~484, and by
the BMBF Grant No.~13N6918A. The author thanks U.~Eckern, J.~Mannhart,
Yu.~Barash, T.~Kopp for useful discussions.

\newpage


\newpage


\begin{references}
\bibitem{domains} The stabilization of the phase ($-$) requires equal charge densities on the
both boundaries of a FE-layer and thus is expected to be suppressed in the
FE-S-FE sandwiches.
\end{references}
\end{document}